\documentclass[10pt, conference, compsocconf]{IEEEtran}

\usepackage{balance}  
\usepackage{graphics} 
\usepackage{graphicx}
\usepackage{epstopdf}
\usepackage{algorithm}
\usepackage{algorithm2e}
\usepackage{times}    
\usepackage{url}      
\usepackage{algorithmic}
\usepackage{amsmath}

\begin{document}

\title{Distributed Bayesian Probabilistic\\ Matrix Factorization}

\author{\IEEEauthorblockN{Tom Vander Aa and Imen Chakroun}
           \IEEEauthorblockA{ExaScience Life Lab, Belgium\\
           IMEC, Leuven, Belgium}

           \and
           
           \IEEEauthorblockN{Tom Haber}
           \IEEEauthorblockA{ExaScience Life Lab, Belgium\\
           UHasselt, Belgium}}
                      
\maketitle

\begin{abstract}
Matrix factorization is a common machine learning technique for recommender
systems. Despite its high prediction accuracy, the Bayesian Probabilistic
Matrix Factorization algorithm (BPMF) has not been widely used on large scale
data because of its high computational cost. In this paper we propose a
distributed high-performance parallel implementation of BPMF on shared memory and
distributed architectures. We show by using efficient load balancing using work
stealing on a single node, and by using asynchronous communication in the
distributed version we beat state of the art implementations.
\end{abstract}

\

\section{Introduction}
\label{introduction}

Recommender Systems (RS) have become very common in recent years and are useful
in various real-life applications.

The most popular ones are probably suggestions for movies on Netflix and books
for Amazon. However, they can also be used in more unlikely area such drug
discovery where a key problem is the identification of candidate molecules that
affect proteins associated with diseases. One of the approaches that have been
widely used for the design of recommender systems is collaborative filtering
(CF). This approach analyses a large amount of information on some users'
preferences and tries to predict what other users may like. A key advantage of
using collaborative filtering for the recommendation systems is its capability
of accurately recommending complex items (movies, books, music, etc) without
having to understand their meaning.  For the rest of the paper, we refer to the
items of a recommender system by movie and user though they may refer to
different actors (compound and protein target for the ChEMBL benchmark for
example \cite{ChEMBL}).

\begin{figure}[h!]
\includegraphics[width=\columnwidth]{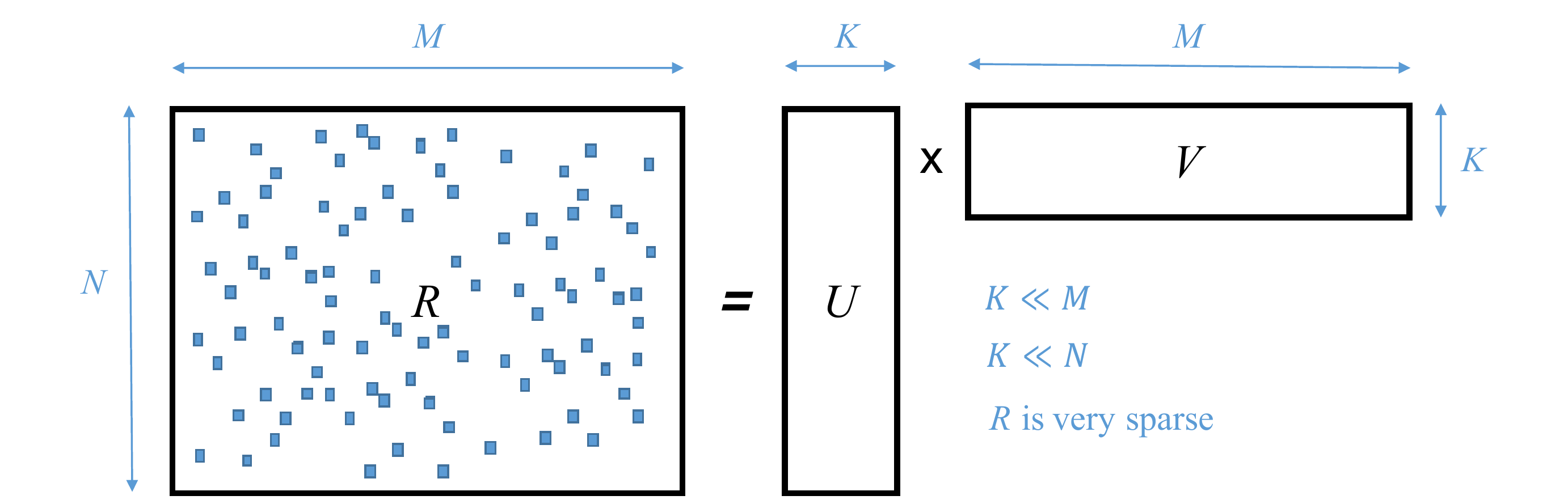}
\caption{Low-rank Matrix Factorization}
\label{fig:mf}
\end{figure}

To deal with collaborative filtering challenges such as the size and the
sparseness of the data to analyze, Matrix Factorization (MF) techniques have
been successfully used. Indeed, they are usually more effective because they
take into consideration the factors underlying the interactions between users
and movies called \emph{latent features}. As sketched in Figure~\ref{fig:mf},
the idea of these methods is to approximate the user-movie rating matrix $R$ as
a product of two low-rank matrices $U$ and $V$ (for the rest of the paper $U$
refers to the users matrix and $V$ to the movie matrix) such that $R \approx
U \times V$. In this way $U$ and $V$ are constructed from the known ratings in
$R$, which is usually very sparsely filled. The recommendations can be made from
the approximation $U \times V$ which is dense.  If $M$ $\times$ $N$ is the
dimension of $R$ then $U$ and $V$ will have dimensions $M$ $\times$ $K$ and $N$
$\times$ $K$.  $K$ represents then number of latent features characterizing the
factors, $K \ll M$, $K \ll N$.

Popular algorithms for low-rank matrix factorization are alternating
least-squares (ALS) \cite{parallelALS}, stochastic gradient descent (SGD)
\cite{parallelSGD} and the Bayesian probabilistic matrix factorization (BPMF)
\cite{BPMF}. Thanks to the Bayesian approach, BPMF has been proven to be
more robust to data-overfitting and released from cross-validation (needed for
the tuning of regularization parameters). In addition, BPMF easily incorporates
confidence intervals and side-information \cite{SIDEINFORMATION, simm:macau}.
Yet BPMF is more computational intensive and thus more challenging to implement
for large datasets.  Therefore, the contribution of this work is to propose a
parallel implementation of BPMF that is suitable for large-scale distributed
systems. 

The remainder of this paper is organized as follows.  Section~\ref{sec:BPMF}
describes the BPMF algorithm. In Section~\ref{sec:MBPMF}, the shared-memory
version and in Section~\ref{sec:DBPMF} the distributed version of BPMF are
described.  The experimental validation and associated results are presented in
Section~\ref{sec:experiments}.  Conclusions  are drawn in
Section~\ref{sec:conclusion} 

\section{BPMF} \label{sec:BPMF}


The BPMF algorithm \cite{BPMF} puts matrix factorization in a Bayesian
framework by assuming a generative probabilistic model for ratings with prior
distributions over parameters.  It introduces common multivariate Gaussian
priors for each user of $U$ and movie in $V$. To infer these two priors from the
data, BPMF places fixed uninformative Normal-Wishart hyperpriors on them. We use
a Gibbs sampler to sample from the prior and hyperprior distributions.

This sampling algorithm can be expressed as the pseudo code shown in
Algorithm~\ref{algo:bpmf_pseudo}. Most time is spent in the loops updating $U$
and $V$, where each iteration consist of some relatively basic matrix and
vector operations on $K \times K$ matrices, and one computationally more
expensive $K \times K$ matrix inversion.

\begin{algorithm}
  \LinesNumbered
  \For{sampling iterations}
  {
      sample hyper-parameters movies based on V

      \For{all movies $m$ of $M$}
      {
          update movie model $m$ based on
          ratings ($R$) for this movie and 
          model of users that rated this movie,
          plus randomly sampled noise
      }

      sample hyper-parameters users based on U

      \For{all users $u$ of $U$}
      {
          update user $u$ based on
          ratings ($R$) for this user and 
          model of movies this user rated,
          plus randomly sampled noise
      }

      \For{all test points}
      {
          predict rating and compute RMSE
      }

  }
  \caption{BPMF Pseudo Code} 
  \label{algo:bpmf_pseudo}
\end{algorithm}%

These matrix and vector operations are very well supported in
Eigen~\cite{Eigen} a high-performance modern C++11 linear algebra library.
Sampling from the basic distributions is available in the C++ standard template
library (STL), or can be trivially implemented on top. As a results the
Eigen-based C++ version of Algorithm~\ref{algo:bpmf_pseudo} is a mere 35 lines
of C++ code with good performance. 

In the next sections we describe how to optimize this implementation to run
efficiently on a shared memory multi-core system and on a distributed
system with multiple compute nodes.


\section{Multi-core BPMF}
\label{sec:MBPMF}

The main challenges for performing BPMF in parallel is how to distribute the
data and the computations amongst parallel workers (threads and/or distributed
nodes).  For the shared memory architectures, our main concern is using as
many cores as possible, keeping all threads as busy as possible and minimizing
memory discontinuous accesses.  Since the number of users entries (resp. movie
entries) is very large and since all items can be computed in parallel, it makes
sense to assigned a set of items to each thread. 

\begin{figure}
\includegraphics[width=\columnwidth]{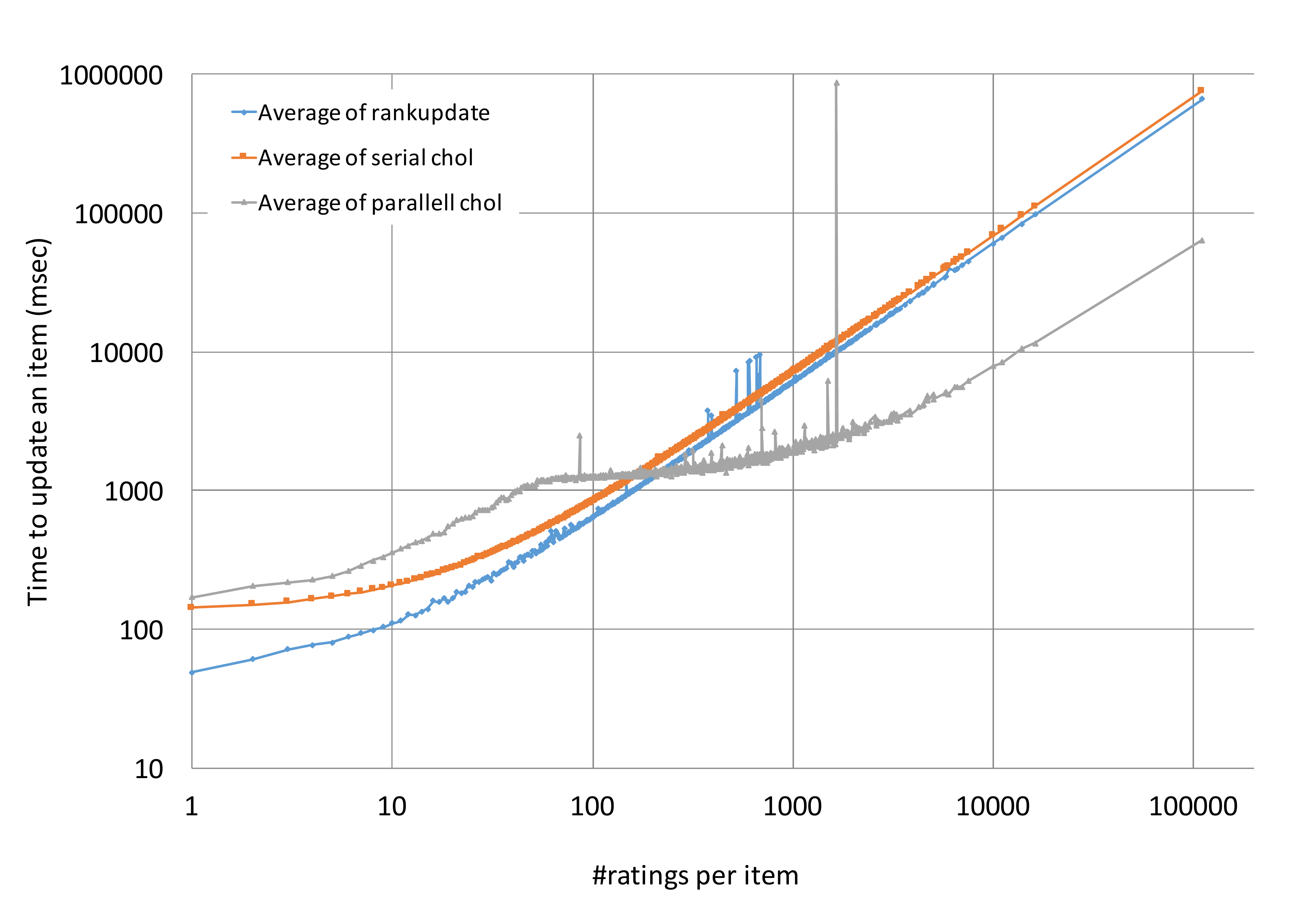}
\caption{Compute time to update one item for the three methods: sequential
rank-one update, sequential Cholesky decomposition, and parallel Cholesky
decomposition}
\label{fig:parallel_sample}
\end{figure}

Next, balanced work sharing is a major way of avoiding idle parallel threads.
Indeed, if the amount of computations is not balanced some threads are likely
to finish their tasks and stay idle waiting for others to finish.  Some items
(users or movies) have a large number of ratings and the amount of compute is
substantially larger for those items. To ensure a good load balance, we use a
cheaper but serial algorithm for items with less than 1000 ratings. For items
with more ratings, we use a parallel algorithm containing a full Cholesky
decomposition. This choice is motivate by Figure~\ref{fig:parallel_sample}
which shows the time to update one item versus the number of ratings for the
three possible algorithms.  By using the parallel algorithm for more expensive
users/movies we effectively split them up in more smaller tasks that can
utilize multiple cores on the system.

\section{Distributed parallel BPMF} \label{sec:DBPMF}

The multi-core BPMF implementation presented above has been extended to
distributed systems using MPI~\cite{MPI}. In this section we
first describe the MPI programming model, next how the data is distributed
across nodes, how the work per node is balanced and how communication is
handled.

\subsection{Distributed Programming using MPI}

Message Passing Interface (MPI) is a standardized and portable message-passing
system for distributed systems. The latest standard MPI-3.0 includes
features important for this BPMF implementation, for example: support for
asynchronous communication, support for hybrid application combining message
passing with shared memory level parallelism like OpenMP~\cite{openmp} or TBB~\cite{tbb}.

\subsection{Data Distribution}

We distribute the matrices $U$ and $V$ across the system where each nodes
computes their part.  When an item is computed, the rating matrix $R$ determines
to what nodes this item needs to be sent.

Our main optimization concern on how to distribute $U$ and $V$ is to make sure
the computational load is distributed as equally as possible and the amount of data
communication is minimized.  Similarly to the cache optimization mentioned
above, we can reorder the rows and columns in $R$ to minimize the number of
items that have to be exchanged, if we split and distribute $U$ and $V$
according to consecutive regions in $R$.

Additionally we take work balance in to account when reordering $R$. For this
we use a workload model derived from Figure~\ref{fig:parallel_sample}: we
approximate the workload per user/movie with fixed cost, plus a cost per movie
rating.

\subsection{Updates and data communication}

To allow for communication and computation to overlap we send the updated
user/movie as soon as it has been computed. For this we use the asynchronous MPI
3.0 routines \texttt{MPI\_Isend} and \texttt{MPI\_Irecv}. However, the overhead
of calling these routines is too much to individually send each item to the
nodes that need it. Additionally, too many messages would be in flight at the
same time for the runtime to handle this efficiently. Hence we store items that
need to be sent in a temporary buffer and only send when the buffer is full.

\section{Validation} \label{sec:experiments}

In this section, we present the experimental results and related discussion for
the proposed parallel implementations of the BPMF described above. 

\subsection{Hardware platform} 

We performed experiments on Lynx a cluster with 20 nodes, each equipped with
dual 6-core Intel(R) Westmere CPUs with 12 hardware threads each, a clock speed
2.80GHz and 96 GB of RAM, and on Fermi, an IBM BlueGene/Q system with 10240
nodes, each equipped with 16 cores running at 1.2Ghz and 16 GB of memory.

\subsection{Benchmarks} \label{bench}

Two public benchmarks have been used to evaluate the performances of the
proposed approaches: the ChEMBL dataset \cite{ChEMBL} and the
MovieLens~\cite{harper:movielens} database. 

The ChEMBL dataset is related to the drug discovery research field. It contains
descriptions for biological activities involving over a million chemical
entities, extracted primarily from scientific literature. Several version exist
since the dataset is updated on a fairly frequent basis. In this work, we used
a subset of the version 20 of the database which was released on February 2015.
The subset is selected based on the half maximal inhibitory concentration
(IC50) which is a measure of the effectiveness of a substance in inhibiting a
specific biological or biochemical function. The total ratings number is around
1023952 from 483500 compounds (acting as users) and 5775 targets (acting as
movies). 

The MovieLens dataset (ml-20m) describes 5-star rating and free-text tagging
activity from MovieLens, a movie recommendation service.  It contains 20M
ratings across 27278 movies. These data were created by 138493 users between
January 09, 1995 and March 31, 2015.

For all the experiments, all the versions of the parallel BPMF reach the same
level of prediction accuracy evaluated using the root mean square error metric
(RMSE) which is a used measure of the differences between values predicted by a
model or an estimator and the values actually observed \cite{Hyndman:RMSE}.

\subsection{Results for Multi-core BPMF}

In this section, we compare the performance of the proposed multi-core
BPMF with the Graphlab library which is a state of the art library widely used
in machine learning community. We have chosen GraphLab because it is known to
outperform other similar graph processing implementations~\cite{Guo:graphlab}.

\begin{figure}
\includegraphics[width=\columnwidth]{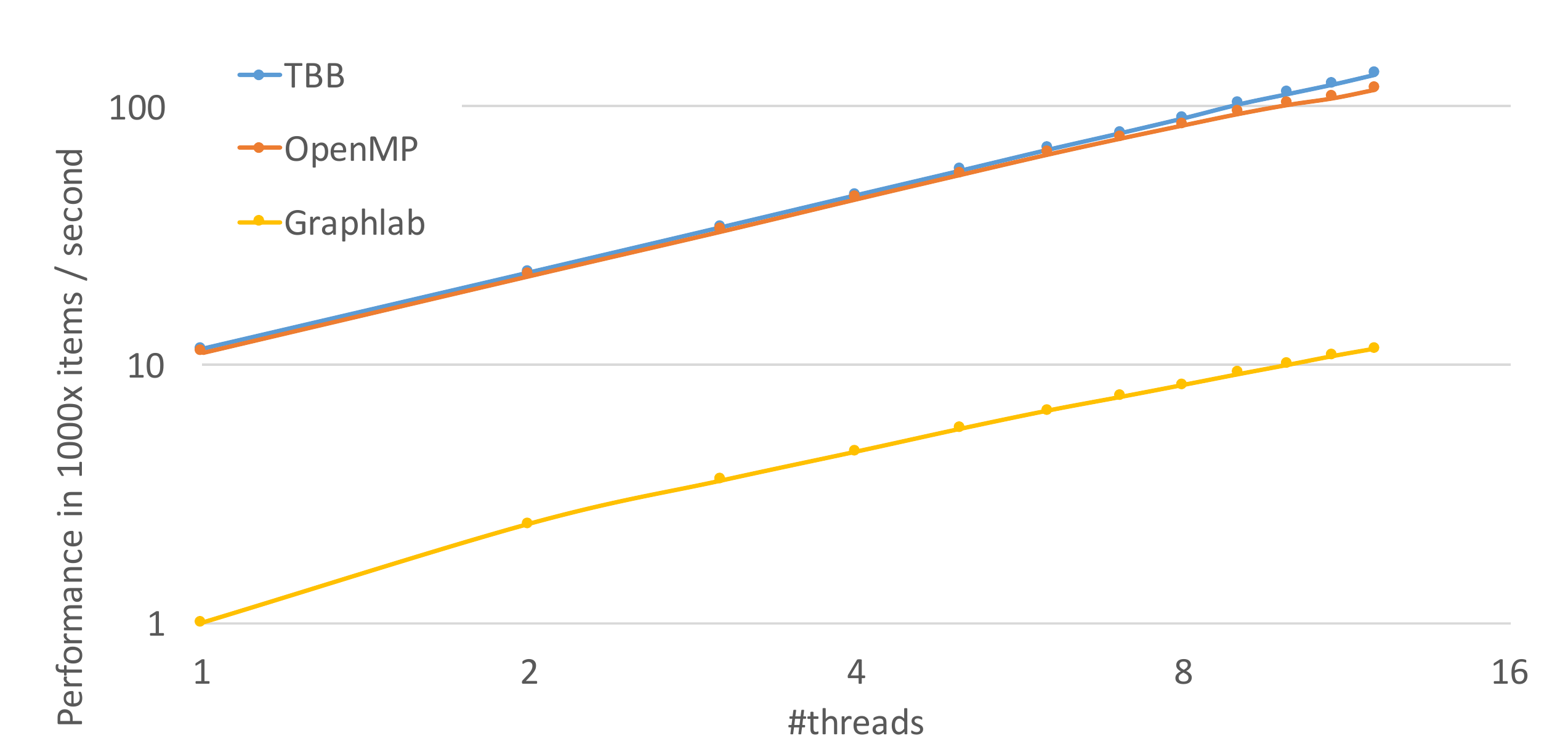}
\caption{Performance of the multi-core BPMF on the ChEMBL dataset in number of
updates to $U$ and $V$ versus the number of parallel threads.}
\label{fig:multicore}
\end{figure}

The results presented in Figure \ref{fig:multicore} report the performance in
number of updates to $U$ and $V$ per second for the ChEMBL benchmark suite on a
machine with 12 cores for three different version, using TBB, OpenMP and GraphLab.

The results show that all parallel implementations of the BPMF scale with the
increasing number of used cores.  The TBB version performs better than the
OpenMP version because TBB's support for nested parallelism and because TBB
uses a work-stealing scheduler that can better balance the work.  The
higher-level GraphLab library focuses less on performance and more on
programmer productivity and this gap is clearly visible in the graph. 

\subsection{Distributed BPMF} \label{Distributed BPMF}

%

\begin{figure*}
    \centering
    \includegraphics[width=0.75\textwidth]{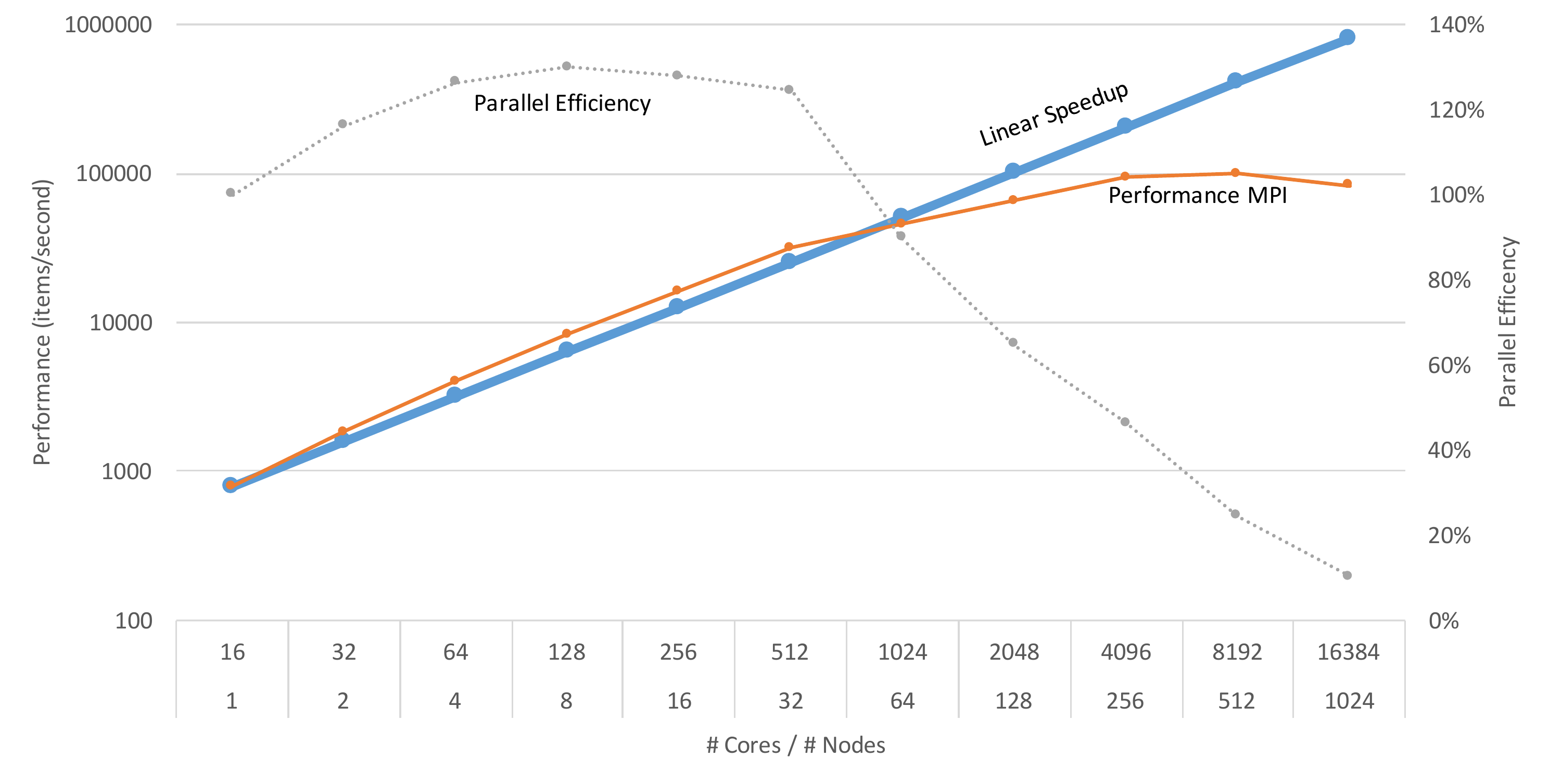}
\caption{Performance of the distributed BPMF on the MovieLens dataset in number of
updates to $U$ and $V$ per second versus the number of cores used. }
\label{fig:fermi}
\end{figure*}

\begin{figure}
\centering
\includegraphics[width=\columnwidth]{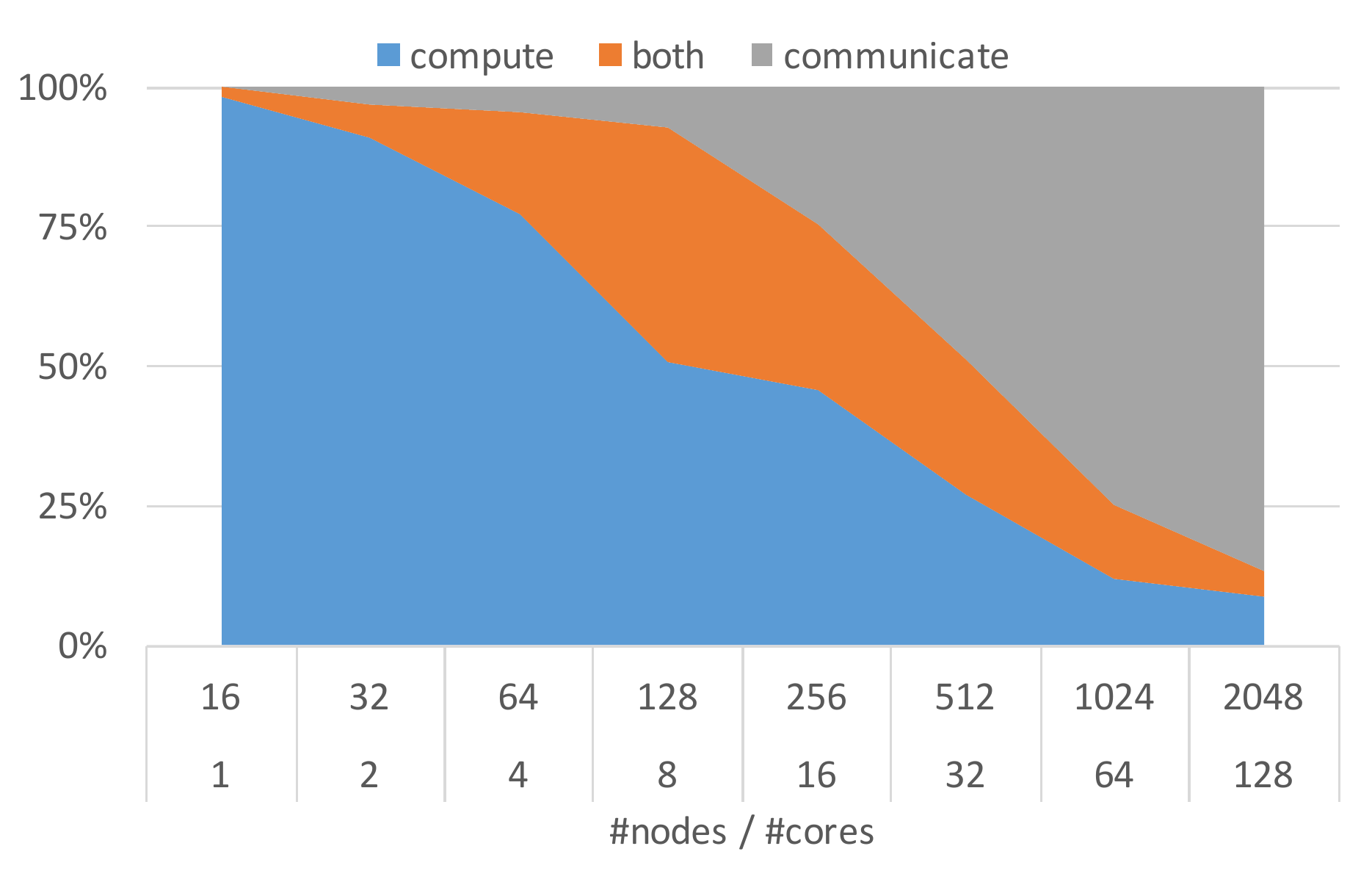}
\caption{Time spent in computing, communication and in both for the distributed
MPI implementation.}
\label{fig:overlap}
\end{figure}

Figure~\ref{fig:fermi} shows strong scaling results for the distributed MPI version
of BPMF on a large system. Scaling is good, even super-linear, up to 32 nodes,
which is one node rack on this system. Once the application's allocation exceeds
this one rack, performance degrades significantly.  The overlap of
communication and computation is displayed in Figure~\ref{fig:overlap}. The term
'both' in this figure means time spent communicating \emph{and} computing. This
figure shows that it is possible to overlap computation and communication with
MPI at small core count. It seems that this overlap does not help much for large
core count, most probably due to a large overhead in the MPI library
itself.

\section{Conclusion and future work}
\label{sec:conclusion}

This work proposed a high-performance distributed implementation of the
Bayesian probabilistic matrix factorization algorithm.  We have shown that load
balancing and asynchronous communication are essential to achieve
good parallel efficiency, clearly outperforming more common synchronous
approaches like GraphLab.  The achieved speed-up allowed us to speed up machine
learning for drug discovery on an industrial dataset from 15 days for the
initial Julia-based version to 30 minutes using the distributed version.

In future work we will try to improve scaling results of the distributed
version of BPMF by using a more light-weight multi-threaded communication
library~\cite{grunewald:gaspi}.


\section*{Acknowledgments}

This work is partly funded by the European projects EXA2CT (EXascale Algorithms and Advanced Computational
Techniques) and ExCAPE (Exascale Compound Activity Prediction Engine) with
references 610741 and 671555.  We acknowledge PRACE for awarding us access to
resource Fermi based in Italy at CINECA and IT4I for providing access to the
Anselm and Salamon systems.

\vfill


\end{document}